\documentclass[runningheads,a4paper]{llncs}

\usepackage{amssymb}
\usepackage{amsmath}
\usepackage{booktabs} 
\usepackage{cite}
\usepackage{graphicx}
\usepackage{url}
\usepackage{xspace}
\usepackage[defblank]{paralist}
\usepackage{epstopdf}
\usepackage{hyphenat}
\usepackage{stmaryrd}
\usepackage{color}
\usepackage{fancyvrb}
\usepackage{tikz}
\usetikzlibrary{patterns}
\usetikzlibrary{backgrounds}
\usepackage{pgfplots}

\usepackage[ruled, linesnumbered]{algorithm2e}

\SetAlFnt{\small}
\SetAlCapFnt{\small}
\SetAlCapNameFnt{\small}
\SetAlCapHSkip{0pt}
\IncMargin{-\parindent}

\usepackage{macros_ld} 

\begin{document}

\mainmatter

\title{Enabling Fine-grained RDF Data Completeness Assessment} 


\author{Fariz Darari$^*$ \and Simon Razniewski \and Radityo Eko Prasojo \and Werner Nutt}

\authorrunning{Fariz Darari \and Simon Razniewski \and Radityo Eko Prasojo \and Werner Nutt}

\institute{Free University of Bozen-Bolzano, Italy\\
	\texttt{$^*$fariz.darari@stud-inf.unibz.it\\}}

\maketitle

\begin{abstract}
Nowadays,
more and more RDF data is becoming available on the Semantic Web.
While the Semantic Web is generally incomplete by nature,
on certain topics, it already contains complete information
and thus, queries may return all answers that exist in reality.
In this paper we develop a technique to check query completeness based on RDF data annotated with completeness information, taking into account data-specific inferences that lead to an inference problem which is $\Pi^P_2$-complete.
We then identify a practically relevant fragment of completeness information,
suitable for crowdsourced, entity-centric RDF data sources such as Wikidata, for which we develop an indexing technique that allows to scale completeness reasoning to Wikidata-scale data sources.
We verify the applicability of our framework using Wikidata
and develop COOL-WD,
a completeness tool for Wikidata, used to annotate Wikidata with completeness statements and
reason about the completeness of query answers over Wikidata.
The tool is available at~\url{http://cool-wd.inf.unibz.it/}.
\end{abstract}



\section{Introduction}
\label{sec:intro}

\begin{figure}[!htbp]
	\centering
	\includegraphics[width=.8\textwidth]{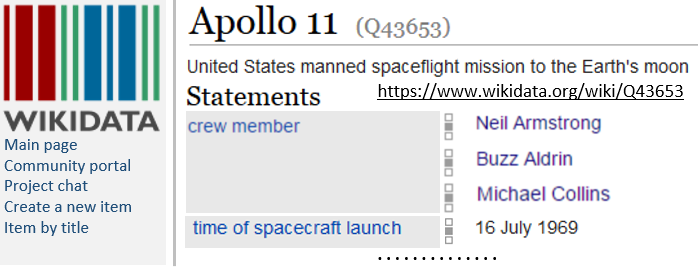}
	\caption{
		Wikidata is actually complete for all the Apollo 11 crew
		}
	\label{fig:apollo}
	\vspace{-2em}
\end{figure}

Over the Web,
we are witnessing a growing amount of data available in RDF.
The LOD Cloud\footnote{\url{http://lod-cloud.net/}} recorded that there were 1014 RDF data sources in 2014,
covering various domains from life science to government.
\newaddition{RDF follows the Open-World Assumption (OWA), assuming data is incomplete by default~\cite{W3C:rdfsemantics}.
Yet,
given such a large quantity of RDF data,
one might wonder if it is complete for some topics.}
As an illustration,
consider Wikidata,
a crowdsourced KB with RDF support~\cite{VrandecicK14}.
For data about the movie Reservoir Dogs,
Wikidata is incomplete,
as it is missing the fact that Michael Sottile was acting in the movie.\footnote{By comparing the data at \url{https://www.wikidata.org/wiki/Q72962} with the complete information at \url{http://www.imdb.com/title/tt0105236/fullcredits}}
On the other hand, for data about Apollo 11,
it is the case that Neil Armstrong, Buzz Aldrin, and Michael Collins, who are recorded as crew members on Wikidata, are indeed \emph{all} the crew (see Figure~\ref{fig:apollo}).\footnote{\url{http://www.space.com/16758-apollo-11-first-moon-landing.html}}
However,
such completeness information is not recorded and thus it is left to the reader to decide
whether some data on the Web is already complete.
	

Nevertheless,
the availability of explicit completeness information can benefit data access over RDF data sources, commonly done via SPARQL queries. To illustrate,
suppose that in addition to the complete data of Apollo 11 crew, Wikidata is also complete for the children of the three astronauts.
Consequently, a user asking
for the children of Apollo 11 crew should obtain not only query answers, but also the information that the answers are complete.

%

Motivated by the above rationales,
we argue that
it is important to describe completeness of RDF data and provide a technique to check query completeness based on RDF data with its completeness information.
Such a check is called \emph{completeness entailment}.
In previous work,
Darari et al.~\cite{DarariNPR13}
proposed a framework to describe completeness of RDF data and check query completeness based on completeness information.
One fundamental limitation of this work is that the completeness check is agnostic of the content of the RDF data to which the completeness information is given, which results in weaker inferences.
In the next section,
we show that incorporating the content of RDF data may provide stronger inferences about query completeness. 
From the relational databases,
Razniewski et al.~\cite{RazniewskiKNS15},
proposed wildcard-based completeness patterns to provide completeness information over databases.
To check query completeness, they defined a pattern algebra, which works upon database tables enriched with completeness patterns.
The work incorporated database instances in completeness check,
which are conceptually similar to the content of RDF data.
However,
only a sound algorithm was provided for completeness check.

In this work, we make the following contributions:
\begin{enumerate}
	\item We provide a formalization of the completeness entailment problem for RDF data,
	and develop a sound and complete algorithm to solve the completeness entailment problem.
	\item We identify a practically relevant fragment of completeness information suitable for crowdsourced, entity-centric RDF data sources like Wikidata, and develop an indexing technique to improve the feasibility of completeness entailment within the fragment. 
	\item We develop COOL-WD, a tool to manage completeness over Wikidata.
\end{enumerate}

Our paper is structured as follows:
Section~\ref{sec:motivation} presents a motivating scenario.
In Section~\ref{sec:formal}, we provide a formalization to the completeness problem, followed by Section~\ref{sec:check} where we describe formal notions 
and a generic algorithm to check completeness entailment.
Section~\ref{sec:sp-statements} introduces a fragment of completeness information, suitable for crowdsourced, entity-centric RDF KBs like Wikidata, 
and presents an optimization technique for checking completeness entailment within this fragment.
Section~\ref{sec:experiments} reports our experimental evaluations.
In Section~\ref{sec:cornerwd}, we describe COOL-WD.
Related work is given in Section~\ref{sec:related}, whereas further discussion about our work is in Section~\ref{sec:discussion}. Section~\ref{sec:conclusions} concludes the paper and sketches future work.

\section{Motivating Scenario}
\label{sec:motivation}


%
%
%

Let us consider a motivating scenario for the main problem of this work, that is, the check of query completeness based on RDF data with its completeness information.
Consider an RDF graph about the crew of Apollo 99 (or for short,~A99), a fictional space mission, and the children of the crew, as shown below.

\begin{figure}[!htbp]
	\vspace{-1.2em}
	\centering
	\includegraphics[width=.5\textwidth]{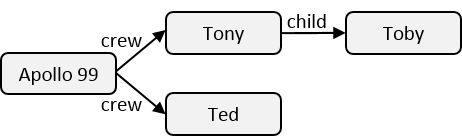}
	\vspace{-1.2em}
\end{figure}

Consider now the query $Q_0$ asking for the crew of A99 and their children:
\[
Q_0 = (W_0, P_0) = (\set{\mathit{?crew}, \mathit{?child}}, \set{\triple{a99}{crew}{?crew}, \triple{?crew}{child}{?child}})
\]
Evaluating $Q_0$ over the graph gives only one mapping result, where the crew is mapped to Tony and the child is mapped to Toby. Up until now, nothing can be said about the completeness of the query since: ($i$) there can be another crew member of A99 with a child; ($ii$) Tony may have another child; or ($iii$) Ted may have a child.

Let us now consider the same graph as before, now enriched with completeness information, as shown below.

\begin{figure}[!htbp]
	\vspace{-1.2em}
	\centering
	\includegraphics[width=.6\textwidth]{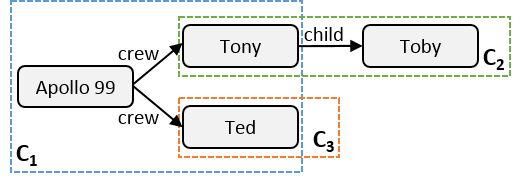}
	\vspace{-1.2em}
\end{figure}

Informally,
the above figure contains three completeness statements:
$C_1$, which states that the graph contains all crew members of A99; $C_2$, which states the graph contains all Tony's children; and $C_3$, which states the graph contains all Ted's children (i.e., Ted has no children).
With the addition of completeness information,
let us see whether we may answer our query completely.

First,
since we know that all the crew of A99 are Tony and Ted, the query~$Q_0$ then becomes equivalent to the following two queries:
\begin{itemize}
	\item $Q_1 = (W_1, P_1) = (\set{?\textit{child}}, \set{\triple{a99}{crew}{tony}, \triple{tony}{child}{?child}})$
	\item $Q_2 = (W_2, P_2) = (\set{?\textit{child}}, \set{\triple{a99}{crew}{ted}, \triple{ted}{child}{?child}})$
\end{itemize}
where the variable $\mathit{?crew}$ is instantiated with Tony and Ted, respectively.

Moreover, for $Q_2$ according to our graph and completeness information, Ted has no children. Thus, there is no way that $Q_2$ will return an answer, so $Q_2$ can be safely removed. Now, only $Q_1$ is left.
Again, from our graph and completeness information,
we know that Toby is the only child of Tony. Thus, $Q_1$ in turn is equivalent to the following boolean query:
\begin{center}
	\vspace{-0.5em}
	$Q_3  = (W_3, P_3) = (\{\}, \set{\triple{a99}{crew}{tony}, \triple{tony}{child}{toby}})$
	\vspace{-0.5em}
\end{center}
with the variable $\mathit{?crew}$ is instantiated to Tony and $\mathit{?child}$ to Toby.
However,
our graph is complete for $Q_3$
as it contains the whole body of $Q_3$.
Since from our reasoning
the query $Q_3$ is equivalent to our original query $Q_0$,
we conclude that our graph with its completeness information can guarantee the completeness of~$Q_0$, that is,
Toby is the only child of Tony, the only crew member of A99 having a child.

Note that using the data-agnostic approach from~\cite{DarariNPR13},
it is not possible to derive the same conclusion.
Without looking at the actual graph, we cannot conclude that Ted and Tony are all the crew members of Apollo 99.
Consequently, just having the children of Tony and Ted complete does not help in reasoning about Apollo 99.
%
In the rest of the paper,
we discuss how the intuitive, data-specific reasoning from above can be formalized.

\vspace{-1em}
\section{Formal Framework}
\label{sec:formal}
\vspace{-0.5em}

In this section,
we remind the reader of RDF and SPARQL, and provide formalization to our completeness problem.

\paragraph*{RDF and SPARQL.}

Assume three pairwise disjoint
infinite sets $I$
(\emph{IRIs}),
$L$ (\emph{literals}), and $V$~(\emph{variables}).
A tuple $(s, p, o) \in I \times I \times (I \cup L)$ is called a triple, while
a finite set of triples is called an RDF graph.

The standard query language for RDF graphs is SPARQL~\cite{W3C:sparqlasbook}.
At the core of SPARQL lies triple patterns,
which are like triples, but also variables are allowed in each position.
In this work,
we focus on the conjunctive fragment of SPARQL, which uses sets of triple patterns, called  basic graph patterns (BGPs).
Evaluating a BGP $P$ over $G$ gives the set of mappings $\eval{P}{G} = \set{\mu \mid \mu P \subseteq G\mbox{ and } \dom{\mu} = \var{P}}$.
Over $P$,
we define \emph{a freeze mapping} $\freeze$ that maps each variable $\mathit{?v}$ in~$P$ to a fresh IRI $\tilde v$.
From such a mapping,
we construct \emph{the prototypical graph} $\frozen P := \freeze \, P$ that encodes any possible graph that can satisfy the BGP.
A query $Q = (W, P)$ projects the evaluation results of a BGP $P$ to a set $W$ of variables.
Moreover, a \CONSTRUCT query has the abstract form $\defineCONSTRUCT{P_1}{P_2}$ where both $P_1$ and $P_2$ are BGPs. Evaluating a \CONSTRUCT query over $G$ results in a graph where $P_1$ is instantiated with all the mappings in $\eval{P_2}{G}$.

\paragraph{Completeness Statements.}

A completeness statement describes which parts of an RDF graph are complete.
We adopt the definition of completeness statements in~\cite{DarariNPR13}.

\begin{definition}[Completeness Statement]
	A completeness statement $C$ is defined as
	$\sCompl{P_C}$ where $P_C$ is a non-empty BGP. 
\end{definition}

\begin{example}
The completeness statements in our motivating scenario are as follows: $C_1 = \compl{\set{\triple{a99}{crew}{?c}}}$, $C_2 = \compl{\set{\triple{tony}{child}{?c}}}$, and $C_3 = \compl{\set{\triple{ted}{child}{?c}}}$.
%
\end{example}

To serialize completeness statements in RDF,
we refer the reader to~\cite{DarariNPR13}.
Now, let us define the semantics of completeness statements.
First, we associate the \CONSTRUCT query $Q_C = \defineCONSTRUCT{P_C}{P_C}$ to each statement $C$.
From now on,
we fix a graph $G$ upon which we describe its completeness.
Given a graph $G' \supseteq G$, we call $(G, G')$ \emph{an extension pair}.
In general, with no completeness statement,
every extension pair is a valid extension pair,
that is,
$G'$ is a possible state of the ideal world where all the information is complete.
For instance,
without completeness statement,
in the motivating scenario,
all of the following would be valid extensions: That there are more crew members of A99; that Tony has more children; and that Ted has children.
Completeness statements restrict the valid extensions of a graph.


\begin{definition}[Valid Extension Pairs]
	Let $G$ be a graph and $C$ a completeness statement.
	We say that an extension pair $(G, G')$ is valid wrt.\ $C$, written $(G, G') \models C$, if $\eval{Q_C}{G'} \subseteq G$.
\end{definition}

The above definition naturally extends to sets of completeness statements.
Over a set $\C$ of completeness statements and a graph $G$, we define \emph{the transfer operator} $T_\C(G) = \bigcup_{C \in \C} \eval{Q_C}{G}$. 
We have the following characterization:
for all extension pairs $(G, G')$,
it is the case that $(G, G')\models \C$
iff
$T_{\C}(G') \subseteq G$. 

\paragraph{Query Completeness.}

We write $\compl{Q}$ to denote that a query $Q$ is complete.
Over an extension pair,
a query is complete iff it returns the same results over both the graphs of the extension pair.


\begin{definition}[Query Completeness]
	Let $(G, G')$ be an extension pair and $Q$ be a query. We define that:
		$(G, G') \models \compl{Q}$ iff $\eval{Q}{G'} = \eval{Q}{G}$.\footnote{Since in this work we focus on conjunctive queries which are monotonic, the direction $\eval{Q}{G'} \supseteq \eval{Q}{G}$ comes for free.}
\end{definition}

\paragraph*{Completeness Entailment}

We now define the main problem of our work, the completeness entailment.


\begin{definition}[Completeness Entailment]
	Given a set $\C$ of completeness statements,
	a graph $G$, and
	a query $Q$,
	we define that
	\emph{$\C$ and $G$ entail the completeness of $Q$},
	written as $\C, G \models \qCompl{Q}$,
	if for all extension pairs $(G, G') \models \C$,
	it holds that $(G, G') \models \qCompl{Q}$.
\end{definition}

In our motivating scenario,
we have seen that the graph about the crew of A99
and the completeness statements there entail
the completeness of the query~$Q_0$ asking for the crew of A99 and their children.

In this work,
we assume bag semantics for query evaluation,
which is the default of SPARQL.\footnote{\url{http://www.w3.org/TR/sparql11-query/}}
Consequently,
this allows us to focus on the BGPs used in the body of conjunctive queries for completeness entailment. 

\vspace{-1.1em} 
\section{Checking Completeness Entailment}
\label{sec:check}

In this section,
we present an algorithm for performing the completeness check as demonstrated in our motivating scenario.

\vspace{-0.4em}
\subsection{Preliminaries}

Before presenting the algorithm,
we introduce important notions.

First, we need to have a notion for a BGP with a stored mapping from variable instantiations.
Let $P$ be a BGP and $\mu$ be a mapping such that
$\dom{\mu}\cap\var{P} = \emptyset$. 
We define the pair $(P, \mu)$ as a \emph{partially mapped BGP}, 
which is a BGP with a stored mapping.
Over a graph $G$, the evaluation of $(P, \mu)$ is defined as
$\eval{(P, \mu)}{G} = \set{ \mu \cup \nu \mid \nu \in \eval{P}{G}}$.
It is easy to see that $P \equiv (P, \emptyset)$.
Furthermore, we define
the evaluation of a set of partially mapped BGPs over a graph $G$ as the union of evaluating each of them over $G$.

\begin{example}
	Consider our motivating scenario.
	Over the BGP $P_0$ of the query~$Q_0$,
	instantiating the variable $\mathit{?crew}$ to $\mathit{\mathit{tony}}$ results in the BGP $P_1$ of the query $Q_1$.
	Pairing $P_1$ with this instantiation gives the partially mapped BGP $(P_1, \set{\mathit{?crew} \mapsto \tony})$.
	Moreover, it is the case that
	$\eval{(P_1, \set{\mathit{?crew} \mapsto \tony})}{G} = \set{\set{\mathit{?crew} \mapsto \tony, \mathit{?child} \mapsto \textit{toby}}}$.
\end{example}


Next,
we would like to formalize the equivalence between partially mapped BGPs wrt.\ a set $\C$ of completeness statements and a graph $G$.

\begin{definition}[Equivalence under $\C$ and $G$]
	Let $(P, \mu)$ and $(P', \nu)$ be partially mapped BGPs,
	$\C$ be a set of completeness statements,
	and $G$ be a graph.
	We define that $(P, \mu)$ is \emph{equivalent} to $(P', \nu)$ wrt.\ $\C$ and $G$,
	written $(P, \mu) \equiv_{\C, G} (P', \nu)$,
	if for all $(G, G') \models \C$, it holds that $\eval{(P, \mu)}{G'} = \eval{(P', \nu)}{G'}$.
	
\end{definition}

The above definition naturally extends to sets of partially mapped BGPs.
\begin{example}
	Consider all the queries in our motivating scenario.
	It is the case that
	$\set{(P_0, \emptyset)} \equiv_{\C, G} \set{(P_1, \set{\mathit{?crew} \mapsto \tony}), (P_2, \set{\mathit{?crew} \mapsto \ted})} \equiv_{\C, G}$
	
	\noindent
	$\set{(P_3, \set{\mathit{?crew} \mapsto \tony, \mathit{?child} \mapsto \textit{toby}})}$.
\end{example}

Next, we would like to figure out which parts of a query contain variables that can be instantiated completely.
For this reason,
we define 
\[
\cruc{P}{\C}{G} = P \cap \frozenIDinv (T_{\C}(\tilde P \cup G))
\]
as \emph{the crucial part of $P$ wrt.\ $\C$ and $G$}.
It is the case that $\C, G \models \qCompl{\cruc{P}{\C}{G}}$, that is,
we are complete for the crucial part.
\newaddition{Later on,
we will see that the crucial part is used to `guide' the instantiation process during the completeness entailment check.}
%


\begin{example}
	Consider the query $Q_0 = (W_0, P_0)$ in our motivating scenario.
	We have that
	$\cruc{P_0}{\C}{G} = P_0 \cap \frozenIDinv (T_{\C}(\tilde P_0 \cup G)) = \set{\triple{a99}{crew}{?crew}}$ with $\frozenID = \set{\mathit{?crew} \mapsto \widetilde{\mathit{crew}}, \mathit{?child} \mapsto \widetilde{\mathit{child}}}$.
	\newaddition{Consequently, we can have a complete instantiation of the crew of A99.}
\end{example}

%
%
%
%

The operator below implements the instantiations of a partially mapped BGP wrt.\ its crucial part.

\begin{definition}[Equivalent Partial Grounding]
	Let $\C$ be a set of completeness statements,
	$G$ be a graph,
	and $(P, \nu)$ be a partially mapped BGP.
	We define the operator \emph{equivalent partial grounding:}
	\begin{center}
		$\maxPartialGrounding{(P, \nu)}{\C}{G} = \set{ (\mu P, \nu \cup \mu) \mid \mu \in \eval{\cruc{P}{\C}{G}}{G}}$.
	\end{center}
\end{definition}

%


The following lemma shows that such instantiations produce a set of partially mapped BGPs equivalent to the original partially mapped BGP, hence the name equivalent partial grounding.
\newaddition{The lemma holds since the instantiation is done over the crucial part,
which is complete wrt.\ $\C$ and $G$.}

\begin{lemma}[Equivalent Partial Grounding]
	\label{lem:equivalentPartialGrounding}
	Let $\C$ be a set of completeness statements, $G$ be a graph, and $(P, \nu)$ be a partially mapped BGP.
	We have that
	\begin{center}
		$\set{(P, \nu)}\equiv_{\C, G}\maxPartialGrounding{(P, \nu)}{\C}{G}$.
	\end{center}
\end{lemma}

%
%

\begin{example}
\label{ex:epg}
	Consider our motivating scenario. 
	We have that:
	\begin{itemize}
		\item $\maxPartialGrounding{(P_2, \set{\mathit{?crew} \mapsto \ted})}{\C}{G} = \emptyset$
		\item $\maxPartialGrounding{(P_3, \set{\mathit{?crew} \mapsto \tony, \mathit{?child} \mapsto \textit{toby}})}{\C}{G} = \{(P_3, \{\mathit{?crew} \mapsto \tony,$ 
		
		$\mathit{?child} \mapsto \textit{toby}\})\}$
		\item $\maxPartialGrounding{(P_0, \emptyset)}{\C}{G} = \set{(P_1, \set{\mathit{?crew} \mapsto \tony}), (P_2, \set{\mathit{?crew} \mapsto \ted})}$
	\end{itemize}
\end{example}

Generalizing from the example above,
there are three cases of $\maxPartialGrounding{(P, \nu)}{\C}{G}$:
\begin{itemize}
	\item If $\eval{\cruc{P}{\C}{G}}{G} = \emptyset$, it returns an empty set.
	\item If $\eval{\cruc{P}{\C}{G}}{G} = \set{\emptyset}$, it returns $\{(P, \nu)\}$.
	\item Otherwise, it returns a non-empty set of partially mapped BGPs where some variables in $P$ are instantiated.
\end{itemize}

From these three cases and the finite number of triple patterns with variables of a BGP,
it holds that the repeated applications of the $\maxPartialGroundingSymbol$ operator,
with the first and second cases above as the base cases,
are terminating.
\newaddition{Note that the difference between these two base cases is on the effect of their corresponding \textit{epg} operations,
as illustrated in Example~\ref{ex:epg}: for the first case, the \textit{epg} operation returns an empty set, whereas for the second case, it returns back the input partially mapped BGP.}

We define that a partially mapped BGP $(P, \nu)$ is \emph{saturated} wrt.\ $\C$ and $G$,
if $\maxPartialGrounding{(P, \nu)}{\C}{G} = \set{(P, \nu)},$
that is, if the second case above applies.
Note that the notion of saturation is independent from the mapping in a partially mapped BGP: given a mapping $\nu$, a partially mapped BGP $(P, \nu)$ is saturated wrt.\ $\C$ and~$G$ iff $(P, \nu')$ is saturated wrt.\ $\C$ and~$G$ for any mapping $\nu'$.
Thus,  wrt.\ $\C$ and~$G$ we say that a BGP $P$ is saturated if $(P, \emptyset)$ is saturated.

The completeness checking of saturated BGPs is straightforward as we only need to check if they are contained in the graph $G$.

\begin{proposition}[Completeness Entailment of Saturated BGPs]
	\label{prop:saturationContainment}
	Let $P$ be a BGP, $\C$ be a set of completeness statements, and $G$ be a graph.
Suppose $P$ is saturated wrt.\ $\C$ and $G$. Then, it is the case that: 
		$\C, G \models \qCompl{P}$ iff  $\tilde P \subseteq G$.

\end{proposition}

Based on the above notions, we are ready to provide an algorithm to check completeness entailment.
The next subsection gives the algorithm.

\subsection{Algorithm for Checking Completeness Entailment}

Now we introduce an algorithm to compute all saturated, equivalent partial grounding results of a BGP wrt.\ $\C$ and $G$.
Following from Proposition~\ref{prop:saturationContainment},
we can then check whether all the resulting saturated BGPs are contained in the graph~$G$ to see if the completeness entailment holds.

%


\begin{algorithm}[H]
	\KwIn{A BGP $P_\original$, a set $\C$ of completeness statements, a graph $G$}
	\KwOut{A set $\Omega$ of mappings}
	\DontPrintSemicolon
	\SetKwFunction{ContainsNonGround}{isNonGround}
	\SetKwFunction{GenEquiv}{genEquiv}
	\SetKwFunction{TakeOne}{takeOne}
	\SetKwFunction{GetMapping}{getMapping}
	
	$\mathbf{P_\working} \gets \set{(P_\original, \emptyset)}$ \;
	$\Omega \gets \emptyset$\;
	\While{$\mathbf{P}_\working \neq \emptyset$}
	{
		$(P, \nu) \gets \TakeOne{$\mathbf{P_\working}$}$\;
		$\mathbf{P_\equivalent} \gets \maxPartialGrounding{(P,\nu)}{\C}{G}$ \;
		\eIf{$\mathbf{P_\equivalent} = \set{(P,\nu)}$}
		{
			$\Omega \gets \Omega \cup \nu$\;
		}
		{$ \mathbf{P_\working} \gets \mathbf{P_\working} \cup \mathbf{P_\equivalent} $\;
		}
	}
	\Return $\Omega$
	\caption{$\saturatedMappingBGPs{P_\original}{\C}{G}$}
	\label{algo:saturation}
\end{algorithm}

Consider a BGP $P_\original$, a set $\C$ of completeness statements, and a graph $G$.
The algorithm works as follows:
First, we transform our original BGP $P_\original$ into its equivalent partially mapped BGP $(P_\original, \emptyset)$ and put it in $\mathbf{P_\working}$.
Then, in each iteration of the while loop, we take and remove a partially mapped BGP $(P, \nu)$ from $\mathbf{P_\working}$ via the method \texttt{takeOne}.
Afterwards, we compute $\maxPartialGrounding{(P,\nu)}{\C}{G}$.
As discussed above there might be three result cases here:
($i$) If $\maxPartialGrounding{(P,\nu)}{\C}{G} = \emptyset$, then simply we remove $(P, \nu)$ and will not consider it anymore in the later iteration; ($ii$) If $\maxPartialGrounding{(P,\nu)}{\C}{G} = \set{(P, \nu)}$, that is, $(P, \nu)$ is saturated, then we collect the mapping $\nu$ to the set $\Omega$; and ($iii$) otherwise, we add to $\mathbf{P_\working}$ a set of partially mapped BGPs instantiated from $(P, \nu)$. We keep iterating until $\mathbf{P}_\working = \emptyset$, and finally return the set $\Omega$.

The following proposition follows from the construction of the above algorithm and Lemma~\ref{lem:equivalentPartialGrounding}.

\begin{proposition}
	Given a BGP $P$, a set $\C$ of completeness statements, and a graph $G$, the following properties hold:
	\begin{itemize}
		
		\item For all $\mu \in \saturatedMappingBGPs{P}{\C}{G}$, it is the case that $\mu P$ is saturated wrt.\ $\C$ and $G$.
		
		\item It holds that $\{(P, \emptyset)\} \equiv_{\C, G} \set{(\mu P, \mu) \mid \mu \in \saturatedMappingBGPs{P}{\C}{G}}$.
	\end{itemize}
	\label{prop:algosat}
\end{proposition}

%
%
%
%
%
%
%

%

From the above proposition,
we can derive the following theorem, which shows
the soundness and completeness of the algorithm to check completeness entailment.

\begin{theorem}[Completeness Entailment Check]
	\label{thm:concreteGraphs}
	Let $P$ be a BGP,
	$\C$ be a set of completeness statements, and
	$G$ be a graph.
	It holds that
	$$ \C, G \models \qCompl{P} \quad \mbox{iff} \quad \mbox{ for all } \mu\in \saturatedMappingBGPs{P}{\C}{G} \; . \; \widetilde{\mu P} \subseteq G.$$ 
	
\end{theorem}

%
%
%


\begin{example}
	Consider our motivating scenario.
	We have that $\saturatedMappingBGPs{P_0}{\C}{G} = \set{ \set{\mathit{?crew} \mapsto \tony, \mathit{?child} \mapsto \textit{toby}} }$.
	It is the case that
	for all $\mu \in \saturatedMappingBGPs{P_0}{\C}{G}$, it holds that $\widetilde{\mu P_0} \subseteq G$.
	Thus, by Theorem~\ref{thm:concreteGraphs}
		the entailment $\C, G \models \qCompl{P_0}$ holds.
\end{example}

By reduction from validity of $\forall \exists$3SAT formula,
one can show that the complexity of the completeness entailment is $\Pi^P_2$-complete.

\begin{corollary}[Complexity of Completeness Check]
	Deciding whether the entailment $\C, G \models \compl{P}$ holds,
	given a set $\C$ of completeness statements,
	a graph $G$,
	and a BGP $P$,
	is $\Pi^P_2$-complete. 
\end{corollary}


\label{subsec:practical-optimizations}

In what follows,
we provide optimization techniques
for the algorithm,
which work for generic cases of completeness entailment.

\paragraph{Early failure detection.}
In our algorithm,
the containment checks for saturated BGPs are done at the end.
Indeed,
if there is a single saturated BGP not contained in the graph,
we cannot guarantee query completeness.
Thus,
instead of having to collect all saturated BGPs and then check the containment later on,
we can improve the performance of the algorithm by performing the containment check
right after the saturation check (Line 6 of the algorithm).
So, as soon as there is a failure in the containment check,
we stop the loop and conclude that the completeness entailment does not hold.


\paragraph{Completeness skip.}
Recall the definition of the operator $\maxPartialGrounding{(P, \nu)}{\C}{G} = \set{ (\mu P, \nu \cup \mu) \mid \mu \in \eval{\cruc{P}{\C}{G}}{G}}$,
which relies on the \texttt{cruc} operator.
Now,
suppose that $\cruc{P}{\C}{G} = P$, that is, we are complete for the whole part of the BGP $P$.
Thus,
we actually do not have to instantiate $P$ in the $\texttt{epg}$ operator,
since we know that the instantiation results are contained in $G$ as the consequence of it being complete wrt.\ $\C$ and $G$.
In conclusion,
whenever $\cruc{P}{\C}{G} = P$,
we just remove $(P, \nu)$ from $\mathbf{P_\working}$ and thus skip its instantiations.

\medskip

Despite these optimizations,
for
a large number of completeness statements,
the completeness entailment check may take long.
In the next section,
we identify a practically relevant fragment of completeness statements, for which we develop an indexing technique to make the entailment check feasible.


%

\section{A Practical Fragment of Completeness Statements}
\label{sec:sp-statements}

This section identifies \emph{SP-statements},
a fragment of completeness statements possessing several properties
that are suitable to be used in practice.
In the next sections,
we show by experimental evaluations the feasibility of this fragment
with an indexing technique we describe below, and demonstrate a completeness tool for Wikidata using the fragment.

\subsection{SP-Statements}

\emph{An SP-statement} $\compl{\set{(s, p, \mathit{?v})}}$ is a completeness statement with only one triple pattern in the BGP of the statement,
where the subject and the predicate are IRIs, and the object is a variable.\footnote{\newaddition{We do not allow the subject to be a variable as it is not practically reasonable (e.g., complete for all the entities and values of predicate \texttt{child}).}}
In our motivating scenario,
all the completeness statements are in fact SP-statements.
The statements possess the following properties,
which are suitable for practical use:
\begin{itemize}
	\item Having a simple structure,
	completeness statements within this fragment are easy to create and to be read.
	Thus, they are suitable for \emph{crowdsourced} KBs,
	where humans are involved.
	\item An SP-statement denotes the completeness of all the property values of the entity which is the subject of the statement. This fits \emph{entity-centric} KBs like Wikidata, where data is organized into entities (i.e., each entity has its own data page).
	\item Despite their simplicity,
	SP-statements can be used to guarantee the completeness of more complex queries such as queries whose length is greater than one (as illustrated by our motivating scenario). 
	
\end{itemize}


\subsection{SP-Indexing}

We describe here how to optimize completeness entailment check with SP-statements.
Recall our generic algorithm to check completeness entailment:

\noindent
In the \textit{cruc} operator within the \textit{epg} operator (Line 5 of Algorithm \ref{algo:saturation}),
we have to compute $T_\C(\tilde P \cup G)$,
that is, evaluate all \CONSTRUCT queries of the completeness statements in $\C$ over the graph $\tilde P \cup G$.
This may be problematic if there are a large number of completeness statements in $\C$.
Thus, we want to avoid such costly $T_\C$ applications.
Given that completeness statements are of the form SP-statements,
we may instead look for the statements having the same subject and predicate of the triple patterns in the BGP.
The crucial part of the BGP $P$ wrt.\ $\C$ and $G$ are the triple patterns with the matching subject and predicate of the completeness statements.
\vspace{-0.2em}
\begin{proposition}
	Given a BGP $P$, a graph $G$, and a set $\C$ of SP-statements, it is the case that
	\vspace{-0.8em}
		\[\cruc{P}{\C}{G} = \set{(s, p, o) \in P \mid 
			\mbox{there exists a statement }\compl{\set{(s, p, \mathit{?v})}} \in \C}.\]
\end{proposition}

From the above proposition,
to get the crucial part,
we only have to find
an SP-statement with the same subject and predicate for each triple pattern of the BGP.
\newaddition{In practice,
we can facilitate this search using a standard hashmap,
providing constant-time performance, also for other basic operations such as \texttt{add} and \texttt{delete}.}
The hashmap provides a mapping from the concatenation of the subject and the predicate of a statement to the statement itself.
To illustrate,
the hashmap of the completeness statements in our motivating scenario is as follows:
 $\set{\textit{a99-crew} \mapsto C_1, \textit{tony-child} \mapsto C_2, \textit{ted-child} \mapsto C_3}$.

\section{Experimental Evaluation}
\label{sec:experiments}


Now that we have an indexing technique for SP-statements,
we want to see the performance of completeness check.
To do so,
we perform experimental evaluations with a realistic scenario,
where we compare the runtime of completeness entailment when query completeness can be guaranteed (i.e., the success case), completeness entailment when query completeness cannot be guaranteed (i.e., the failure case), and query evaluation.

\paragraph{Experimental Setup.}
Our reasoning algorithm and indexing modules are
implemented in Java using the Apache Jena library.\footnote{\url{https://jena.apache.org/}}
We use Jena-TDB as the triple store of our experiment.
The SP-indexing is implemented using the standard Java \texttt{hashmap},
where the keys are strings,
constructed from the concatenation of the subject and predicate of completeness statements,
and the values are Java objects representing completeness statements.
All experiments are done on a standard laptop with a 2.4 GHz Intel Core i5 and 8 GB of memory.

To perform the experiment,
we need three ingredients: a graph, completeness statements, and queries.
For the graph,
we use the direct-statement fragment of the Wikidata graph, which does not include qualifiers nor references and consists of 100 mio triples.\footnote{\url{http://tools.wmflabs.org/wikidata-exports/rdf/index.php?content=dump_download.php&dump=20151130}}
The completeness statements and queries of this experiment are constructed based on the following pattern queries:
\begin{enumerate}
	\item Give all mothers of mothers of mothers.
	
	$P_1 = \set{\triple{?v}{P25}{?w}, \triple{?w}{P25}{?x}, \triple{?x}{P25}{?y}}$
	
	\item Give the crew of a thing, the astronaut missions of that crew, and the operator of the missions.
	
	$P_2 = \set{\triple{?v}{P1029}{?w}, \triple{?w}{P450}{?x}, \triple{?x}{P137}{?y}}$
	
	\item Give the administrative divisions of a thing, the administrative divisions of those divisions, and their area.
	
	$P_3 = \set{\triple{?v}{P150}{?w}, \triple{?w}{P150}{?x}, \triple{?x}{P2046}{?y}}$
\end{enumerate}



To generate queries,
we simply evaluate each pattern query over the graph,
and instantiate the variable $\mathit{?v}$ of each pattern query
with the corresponding mappings from the evaluation.
We record 5200
queries instantiated from $P_1$,
57 queries from $P_2$, and
475 queries from $P_3$.
Each pattern query has a different average number of query results: the instantiations of $P_1$ give 1 result, those of~$P_2$ give 4 results, and those of $P_3$ give 108 results on average.

To generate completeness statements,
from each generated query,
we iteratively evaluate
each triple pattern from left to right, and construct SP-statements from the instantiated subject and the predicate of the triple patterns.
This way, we guarantee that all the queries can be answered completely.
We generate in total around 1.7 mio statements, with
30072 statements for $P_1$,
484 statements for $P_2$, and
1682263 statements for $P_3$.
Such a large number of completeness statements would make completeness checks without indexing very slow:
Performing just a single application of the $T_\C$ operator with all these statements,
which occurs in the \texttt{cruc} operator of the algorithm without SP-indexing, took about 20 minutes.
Note that in a completeness check, there might be many $T_\C$ applications.

Now we describe how to observe the behavior when queries cannot be guaranteed to be complete, that is, the failure case.
In this case, we drop randomly 20\% of the completeness statements for each pattern query.
To make up the statements we drop,
we add dummy statements with the number equal to the number of dropped statements.
This way, we ensure the same number of completeness statements for both the success and failure case. 

For each query pattern,
we measure the runtime of completeness check for both the success case and the failure case, and then query evaluation for the success case.\footnote{We do not measure query evaluation time for failure case since query evaluation is independent of the completeness of the query.}
We take 40 sample queries for each pattern query,
repeat each run 10 times,
and report the median of these runs.

\paragraph{Experimental Results.}
\begin{figure}[!t]
	\centering
	\begin{tikzpicture}
	\begin{semilogyaxis}[
	ybar,
	enlarge x limits = 0.1,
	legend style = {at = { (0.5, -0.3)}, anchor = north, legend columns = -1},
	ylabel = {Runtime in $\mu$s},
	xlabel = {Pattern Query},
	symbolic x coords={1, 2, 3},
	xtick = data,
	width = 8cm,
	bar width=6,
	height = 5cm,
	ymin = 0,
	legend image code/.code={
		\draw[#1] (0cm,-0.1cm) rectangle (0.1cm,0.2cm);
	},
	]

	\addplot[draw = black, fill = green] 
	coordinates {
		(1, 796.053)
		(2, 5264.917)
		(3, 35130.149)
	};


	\addplot[draw = black, fill = red] 
	coordinates {
		(1, 485.857)
		(2, 903.306)
		(3, 1209.413)
	};
	

	\addplot[draw = black, fill = yellow] 
	coordinates {
		(1, 29.189)
		(2, 76.216)
		(3, 673.290)
	};
	

	\legend{Success Case, Failure Case, Query Evaluation}
	\end{semilogyaxis}
	\end{tikzpicture}
	\caption{Experiment Results of Completeness Entailment
		}
	\label{fig:real-experiment}
	\vspace{-2em}
\end{figure}
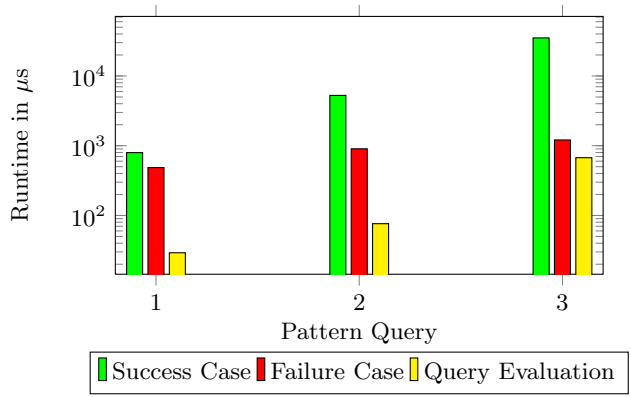
The experimental results are shown in Figure~\ref{fig:real-experiment}.
Note that the runtime is in log scale.
We can see that in all cases,
the runtime increases with the first pattern query having the lowest runtime, and the third pattern query having the highest runtime.
This is likely due to the increased number of query results.
We observe that in all pattern queries,
completeness check when queries are guaranteed to be complete is slower than those whose completeness cannot be guaranteed.
We suspect that this is because in the former case,
variable instantiations have to be performed much more than in the latter case.
In the latter case,
as also described in Subsection~\ref{subsec:practical-optimizations}, as soon as we find a saturated BGP not contained in the graph, we stop the loop in the algorithm and return \texttt{false}, meaning that the query completeness cannot be guaranteed.


In absolute scale, completeness check runs relatively fast, with
796 $\mu$s for~$P_1$, 5264 $\mu$s for $P_2$, and 35130 $\mu$s for $P_3$ in success case;
and 485 $\mu$s for $P_1$, 903 $\mu$s for $P_2$, and 1209 $\mu$s for $P_3$ in failure case.
Note that as mentioned before,
completeness check without indexing is not feasible at all here,
as there are a large number of completeness statements, making the $T_\C$ application very slow (i.e., 20 minutes for a single application).
For all pattern queries, however, query evaluation runs faster than completeness checking.
This is because completeness checking may involve several query evaluations during the instantiation process with the \texttt{epg} operator.

To conclude,
we have observed
that completeness checking with a large number of SP-statements can be done reasonably fast, even for large datasets, by the employment of indexing.
Also,
we observe a clear positive correlation between
the number of query results and the runtime of completeness checking.
Last, performing completeness check when a query is complete is slower than that when a query cannot be guaranteed to be complete.

\section{COOL-WD: A Completeness Tool for Wikidata}
\label{sec:cornerwd}


In this section,
we introduce COOL-WD,
a \textbf{CO}mpleteness to\textbf{OL} for \textbf{W}iki\textbf{D}ata.
The tool implements our completeness framework with SP-statements and focuses to provide completeness information for direct statements of Wikidata.
While our implementation is based on Apache Jena,
our approach can be applied also via other Semantic Web frameworks like Sesame.\footnote{\url{http://rdf4j.org/}}
Our tool is inspired by real, natural language completeness statements on Wikipedia,
where completeness statements are given in a crowdsourced way.\footnote{\url{https://en.wikipedia.org/wiki/Template:Complete_list}}
The tool is available at \url{http://cool-wd.inf.unibz.it/}.

\subsection{System Architecture}

\begin{figure}[t]
	\centering
	\includegraphics[width=0.5\textwidth]{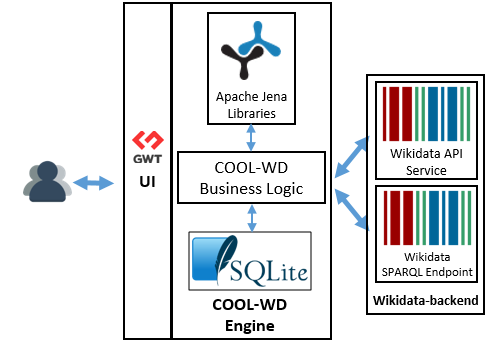}
	\caption{COOL-WD Architecture}
	\label{fig:cornerwd}
\end{figure}

As shown in Figure~\ref{fig:cornerwd}, COOL-WD consists of three main components: user interface (UI), COOL-WD engine, and Wikidata-backend.

The first component is the UI, developed using GWT.\footnote{\url{http://www.gwtproject.org/}}
The UI provides the front-end interface for COOL-WD users, enabling them to search for Wikidata entities, look at facts about them enriched with completeness information, add/remove completeness statements, and check the completeness of a Wikidata query.

The second component is the engine, responsible for storing completeness statements using SQLite and performing completeness checks.
We use optimization techniques as described in Subsection~\ref{subsec:practical-optimizations} and SP-indexing as described in Section~\ref{sec:sp-statements} to improve the performance of completeness checks.

The last component is the Wikidata-backend.
It consists of
two subcomponents: Wikidata API and Wikidata SPARQL endpoint.
The API is used for the suggestions feature in searching for Wikidata entities, while the Wikidata SPARQL endpoint serves as the source of Wikidata facts to which completeness statements are given, and of query evaluation.

%
%

\subsection{Tool Usage}


\begin{figure}[t]
	\centering
	\includegraphics[width=\textwidth]{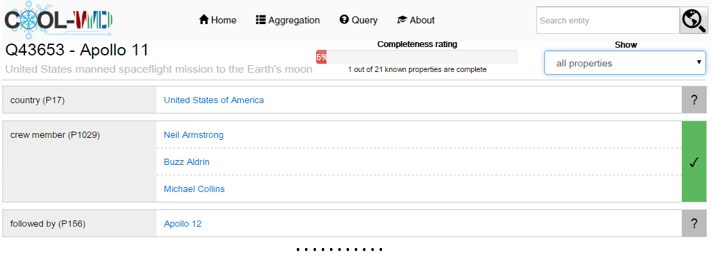}
	\caption{The COOL-WD page of Apollo 11. Complete property values are with checkmarks.
		}
	\label{fig:demo-description}
\end{figure}

Here,
we describe how one can use COOL-WD.
From the landing page, the user is provided with a search bar
for Wikidata entities.
The search bar features auto-complete search suggestions,
matching user keywords with the English labels of Wikidata entities.
Clicking on a search suggestion gives the users the entity page, consisting of Wikidata facts about the entity with its completeness information.
An example is shown in Figure~\ref{fig:demo-description},
which is the Apollo 11 page with the complete crew.
Complete properties are distinguished by the checkmark symbol.
To add a completeness statement,
a user simply clicks a question mark next to the respective properties of an entity.
Additionally, it is possible to add provenance information about authors, timestamps, and references of the statement.
Suppose the user would also like to add completeness statements for the astronaut missions of Neil Armstrong, Buzz Aldrin, and Michael Collins. Therefore, she may perform an analogous operation: go to the entity pages, and click the question mark next to the respective properties. We also have a feature to see all stored completeness statements
over Wikidata filtered by properties on the aggregation page.

If a user would like to evaluate a query and check its completeness,
she has to go to the query page. Suppose she wants to know the crew of Apollo 11 and their astronaut missions.
The user then specifies her query, and executes it.
Instead of having only query answers, she can also see the completeness information of the answers.

\section{Related Work}
\label{sec:related}



Data completeness concerns the breadth,
depth,
and scope of
information~\cite{WangS96}.
In the relational databases,
Motro~\cite{Motro89} and
Levy~\cite{Levy96} were among the first to investigate data completeness.
Motro developed a sound technique to check query completeness based on database views,
while
Levy introduced the notion of local completeness statements to denote which parts of a database are complete.
Razniewski and Nutt~\cite{RazniewskiN11:completeness} further extended their results 
by reducing completeness reasoning to containment checking, for which many algorithms are known,
and characterizing the complexity of reasoning for different classes of queries. In terms of their terminology, our completeness entailment problem is one of QC-QC entailment under bag semantics, for which so far it was only known that it is in $\Pi^P_3$~\cite{RazniewskiNVLDBJsubmission}.
In~\cite{RazniewskiKNS15}, Razniewski et al.\ proposed completeness patterns and defined a pattern algebra to check the completeness of queries.
The work incorporated database instances, yet provided only a sound algorithm for completeness check.

We now move on to the Semantic Web.
F\"urber and Hepp~\cite{FurberH11} distinguished three types of completeness: ontology completeness, concerning which ontology classes and properties are represented;
population completeness,
referring to whether all objects of the real-world are represented; and property completeness,
measuring the missing values of a specific property. In our work, SP-statements can be used to state the property completeness of an entity.
Mendes et al.~\cite{MendesMB12} proposed Sieve, 
a framework for expressing quality assessment and fusion methods, where completeness is also considered.
With Sieve, users can specify how to compute quality scores and express a quality preference specifying which characteristics of data indicate higher quality.
\newaddition{In the context of crowdsourcing,
Chu et al.~\cite{ChuMIOP0Y15} developed KATARA, a hybrid data cleaning system, which not only cleans data, but may also add new facts to increase the completeness of the KB; whereas Acosta et al.~\cite{AcostaSFV15} developed HARE, a hybrid SPARQL engine to enhance answer completeness.}

Gal{\'{a}}rraga et al.\ \cite{GalarragaTHS13} 
proposed a rule mining system that is able to operate under the Open-World Assumption (OWA) by simulating negative
examples using the Partial Completeness Assumption (PCA). The PCA
assumes that if the dataset knows some $r$-attribute of $x$, then it
knows all $r$-attributes of $x$.
\newaddition{This heuristic was also employed by Dong et al.~\cite{DGHHLMSSZ14} to develop Knowledge Vault, a Web-scale system for probabilistic knowledge fusion. In their paper, they used the term Local Closed-World Assumption (LCWA).} 



\section{Discussion}
\label{sec:discussion}

We discuss here various aspects of our work: sources of completeness statements, completeness statements with provenance, and no-value information.

\paragraph{Sources of Completeness Statements.}
As demonstrated by COOL-WD,
one way to provide completeness statements is via crowdsourcing.
For domain-specific data like biology and archeology,
domain experts may be a suitable source of completeness statements.
An automated way to add completeness statements can also be leveraged by using NLP techniques to extract natural language completeness statements already available on the Web: around 13000 Wikipedia pages contain the keywords ``complete list of'' and ``list is complete'', 
while IMDb provides complete cast information with the keywords ``verified as complete'' for some movies like Reservoir Dogs.\footnote{\url{http://www.imdb.com/title/tt0105236/fullcredits}}


\paragraph{Completeness Statements with Provenance.}
Just as data can be wrong,
completeness statements can be wrong, too.
Moreover,
as data may change over time,
completeness statements can be out-of-date.
As a possible solution,
one can add provenance information.
Adding information about the author and reference of completeness statements may be useful to check the correctness of the statements,
while attaching timestamps
would provide timeliness information to the statements.

\paragraph{No-Value Information.}
Completeness statements can also be used to represent the non-existence of information.
For example, in our motivating scenario, there is the completeness statement about the children of Ted with no corresponding data in the graph.
In this case,
we basically say that Ted has no children.
As a consequence of having no-value information,
we can be complete for queries despite having the empty answer.
Such a feature is similar to that proposed in~\cite{DarariEN15}. The only difference is that here we need to pair completeness statements with a graph that has no corresponding data captured by the statements, while in that work,
no-value statements are used to directly say that some parts of data do not exist.


\section{Conclusions and Future Work}
\label{sec:conclusions}

The availability of an enormous amount of RDF data calls for better data quality management.
In this work, we focus on the data quality aspect of completeness.
We develop a technique to check query completeness based on RDF data with its completeness information.
To increase the practical benefits of our framework,
we identify a practically relevant fragment of completeness information upon which an indexing can be implemented to optimize completeness check, and develop COOL-WD, a completeness management tool for Wikidata.

For future work,
we would like to investigate
indexing techniques for more general cases.
One challenge here is that how to index the arbitrary structure of completeness statements.
Another plan is to develop a technique to extract completeness statements on the Web. To do so,
we in particular want to detect if a Web page contains completeness statements in natural language, and transform them into RDF-based completeness statements.
Last, we also want to increase the expressivity of queries, say, to also handle negations.
Queries with negations are especially interesting since negation naturally needs complete information to work correctly.


{\small
\section*{Acknowledgments}
\label{sec:ack}

We would like to thank Sebastian Rudolph for his feedback on an earlier version of this paper.
\newaddition{The research was supported by the projects ``CANDy: Completeness-Aware Querying and Navigation on the Web of Data'' and ``TaDaQua - Tangible Data Quality with Object Signatures'' of the Free University of Bozen-Bolzano, and ``MAGIC: Managing Completeness of Data'' of the province of Bozen-Bolzano.}
}

\bibliographystyle{unsrt}
\bibliography{icwe}

\end{document}